\documentclass[preprint]{aastex}
\begin{document}
\title{Supersonic water masers in 30 Doradus}
\author{Jacco Th. van Loon\altaffilmark{1}}
\affil{Institute of Astronomy, Madingley Road, Cambridge CB3 0HA,
       United Kingdom}
\and
\author{Albert A. Zijlstra\altaffilmark{2}}
\affil{UMIST, P.O.Box 88, Manchester M60 1QD, United Kingdom}
\begin{abstract}

We report on extremely high velocity molecular gas, up to $-80$ km s$^{-1}$
relative to the ambient medium, in the giant star-formation complex 30 Doradus
in the Large Magellanic Cloud (LMC), as observed in new 22 GHz H$_2$O
$6_{16}\rightarrow5_{23}$ maser emission spectra obtained with the Mopra radio
telescope. The masers may trace the velocities of protostars, and the observed
morphology and kinematics indicate that current star formation occurs near the
interfaces of colliding stellar-wind blown bubbles. The large space velocities
of the protostars and associated gas could result in efficient mixing of the
LMC. A similar mechanism in the Milky Way could seed the galactic halo with
relatively young stars and gas.

\end{abstract}

\section{Introduction}

The most luminous H {\sc ii} region in the Local Group of galaxies is 30
Doradus (Tarantula nebula, N157: Kennicutt 1984; Walborn 1991) in the Large
Magellanic Cloud (LMC). It is the site of recent star formation on a massive
scale and harbours a rich stellar cluster (NGC2070) of which the core (R136:
Feast et al.\ 1960) has only recently been resolved into many individual stars
(cf.\ Hunter 1999). The H {\sc ii} region is embedded in an extended halo of
expanding gas. In between, the stellar winds from the rapidly evolving massive
stars interact with each other and with the surrounding interstellar medium
(ISM) to create an intricate complexity of shells, filaments and dense knots
(Meaburn 1980, 1984; Wang \& Helfand 1991; Chu \& Kennicutt 1994). The
radiation from these hot stars photo-evaporates the surrounding dusty,
molecular and neutral clouds (Hunter et al.\ 1995; Poglitsch et al.\ 1995;
Scowen et al.\ 1998). It thus appears that the newly born stars destroy their
parent Giant Molecular Cloud (GMC), which might prevent further star
formation. Yet the recent discovery of protostars and structures similar to
Herbig-Haro objects (Jones et al.\ 1986; Walborn \& Blades 1987, 1997; Hyland
et al.\ 1992; Rubio et al.\ 1992, 1998; Walborn et al.\ 1999) provides ample
evidence for the occurrence of current star formation, in spite of --- or
perhaps due to (Dopita 1981; Caulet et al.\ 1982; Dopita et al.\ 1985) --- the
violent interaction between colliding stellar-wind blown bubbles.

The first evidence of ongoing star formation in 30 Doradus was the detection
of H$_2$O maser emission from 0539$-$691 (Whiteoak et al.\ 1983), a region
$\sim1.5^\prime$ NE of R136. H$_2$O masers have been found in three more
regions in the LMC (Scalise \& Braz 1982; Whiteoak et al.\ 1983; Whiteoak \&
Gardner 1986) amongst which is 0540$-$696 in the GMC extending $\sim3$ kpc due
S of 30 Doradus (Israel 1984; Cohen et al.\ 1988; Kutner et al.\ 1997;
Johansson et al.\ 1998). H$_2$O masers occur in dust-enshrouded (Class 0: Shu
et al.\ 1987) Young Stellar Objects (YSOs) and Ultra Compact H {\sc ii}
regions (e.g.\ Knapp \& Morris 1976; Genzel \& Downes 1977; Cesarsky et al.\
1978; Rodriguez et al.\ 1978; Codella et al.\ 1994, 1996), and disappear once
the H {\sc ii} region becomes diffusive after $\sim10^5$ yr (Genzel \& Downes
1979; Codella \& Felli 1995). H$_2$O maser emission arises both in low-mass
and massive YSOs (Haschick et al.\ 1983; Wouterloot \& Walmsley 1986; Tereby
et al.\ 1992; Wilking et al.\ 1994; Codella et al.\ 1995) but is more intense
in massive cores of star formation (Palla et al.\ 1991; Henning et al.\ 1992;
Palagi et al.\ 1993; MacLeod et al.\ 1998). Models (Elitzur et al.\ 1989,
1992) suggest that the H$_2$O masers occur in shocked layers at the neutral
side of photo-dissociation regions (PDRs) around newly formed stars. Masers
are extremely bright emission lines, that can be used both to probe the
kinematics within the circumstellar environment as well as to trace the
systemic velocity of the masing object.

Here we present new 22 GHz data of the known H$_2$O maser source 0539$-$691
and a hitherto unknown H$_2$O maser at $\sim2.5^\prime$ SE of R136. Both show
discrete emission components at large blue-shifted velocities with respect to
the bulk of the ISM in the 30 Doradus region. This places the masers and hence
the star formation sites at the rim of expanding stellar-wind blown bubbles,
providing new evidence for sequential, shock-induced star-formation.

\section{Observations}

The 22 m radio telescope at Mopra, Australia, was used from January 18 to 29,
1999, with the 1.3 cm receiver plus autocorrelator backend to observe the
$6_{16}\rightarrow5_{23}$ rotational transition of ortho-H$_2$O at a rest
frequency of 22.23507985 GHz. The aim was to detect maser emission from
circumstellar envelopes around evolved giants in the Magellanic Clouds (van
Loon et al.\ 2000). While observing IRAS05280$-$6910, however, one of the
background reference beams, pointed at $(\alpha, \delta)_{\rm J2000} =
(5^h39^m03.9^s, -69^\circ08^\prime18^{\prime\prime})$, picked up a signal from
the 30 Doradus complex. This source will be referred to as 0539$-$691B. It is
located only $\sim3.3^\prime$ away from the known H$_2$O maser 0539$-$691
(Whiteoak et al.\ 1983; Whiteoak \& Gardner 1986) at $(\alpha, \delta)_{\rm
J2000} = (5^h38^m49.6^s, -69^\circ04^\prime41^{\prime\prime})$ that we
observed as well. The beam FWHM was $2.7^\prime$, corresponding to $\sim39$ pc
at the distance of the LMC (50 kpc).

The 64 MHz band width and 1024 channels centred at $\sim$22.215 GHz yielded a
velocity coverage of $\sim$860 km s$^{-1}$ at 0.84 km s$^{-1}$ channel$^{-1}$.
The system temperature $T_{\rm sys}\sim115$ K (50 to 65 K from the sky), and
the opacity correction was between $\sim1.3$ and 1.6. The conversion factor
from antenna temperature to flux density was 20 Jy K$^{-1}$, consistent with
the observed noise and flux density for R Dor (48 Jy). The on-source
integration time on 0539$-$691B was 56 minutes. The baselines were constructed
by heavily smoothing the spectra --- excluding the maser peaks --- with a
gaussian of $\sigma=50$ km s$^{-1}$, and these were then subtracted.

\section{Results}

\subsection{Kinematics}

The 22 GHz emission spectrum of 0539$-$691B peaks sharply at $v_{\rm hel}=201$
km s$^{-1}$ (Fig.\ 1, top panel). In 0539$-$691 the main component peaks at
$v_{\rm hel}\sim266$ km s$^{-1}$ (Fig.\ 1, bottom panel) and may be double as
was also observed in the early 1980's (Whiteoak \& Gardner 1986). The
secondary peak in 0539$-$691, at a remarkably low $v_{\rm hel}\sim182$ km
s$^{-1}$, was not known before. The source of this maser emission may be near
the edge of the Mopra beam and hence outside the smaller Parkes beam. Other
spikes in the spectrum of 0539$-$691 may be spurious. Both 0539$-$691 and
0539$-$691B peak at $\sim0.5$ Jy, which is a few times fainter than 0539$-$691
in the early 1980's. This might be due to temporal variability (see, for
instance, Knapp \& Morris 1976; Hunter et al.\ 1994; Persi et al.\ 1994).

The main emission component of 0539$-$691 around $v_{\rm hel}\sim266$ km
s$^{-1}$ has a 21 cm H {\sc i} counterpart (McGee \& Milton 1966; Kim et al.\
1999), which is seen in deep absorption. OH 1665 \& 1667 MHz absorption is
centred at $v_{\rm hel}\sim262$ km s$^{-1}$ (Gardner \& Whiteoak 1985) and
probably associated with the same region. The bulk of the nebular emission
(Clayton 1987; Chu \& Kennicutt 1994; Hunter 1994), Ca {\sc ii} interstellar
absorption (Wayte 1990), $^{12}$CO$(1\rightarrow0)$ emission (Israel et al.\
1986; Johansson et al.\ 1998) and radio recombination lines (Peck et al.\
1997), as well as the velocities of supergiants (Wayte 1990) are within
$\sim20$ km s$^{-1}$ of a mean $v_{\rm hel}\sim270$ km s$^{-1}$. This
indicates a systemic velocity of the 30 Doradus complex.

Strong H$_2$O maser emission peaks within $\sim20$ km s$^{-1}$ from the
systemic velocity of the parent molecular cloud (Genzel et al.\ 1981; Tereby
et al.\ 1992), because the amplification is tangential whilst the velocity
field is mostly radial with respect to the source of photons (Elitzur 1992).
In some cases masers with relative velocities $>100$ km s$^{-1}$ are detected
--- possibly related to fast molecular outflows from the YSOs --- but these
masers are nearly always (much) fainter than the emission near the systemic
velocity (Morris 1976; Genzel \& Downes 1977; Rodriguez et al.\ 1978; Blitz \&
Lada 1979; Downes et al.\ 1979; Genzel et al.\ 1979). It is therefore unlikely
that two out of three bright H$_2$O masers in 30 Doradus would originate in
such jets. Hence the large differences in velocity with respect to the
systemic velocity of the 30 Doradus complex suggest that the H$_2$O masers at
$v_{\rm hel}=182$ and 201 km s$^{-1}$ are not associated with the bulk of the
molecular gas in the 30 Doradus complex, but rather with compact knots or
slabs in the ISM as they are also seen in the nebular (Scowen et al.\ 1998)
and IR emission (Walborn et al.\ 1999); these structures often move
supersonically with respect to the ambient ISM (e.g.\ Chu \& Kennicutt 1994).
The high-velocity H$_2$O masers may thus trace star formation in supersonic
gas.

\subsection{Morphology}

The locations of the masers with respect to the IR emission in the 30 Dor
region are shown in Fig.\ 2 (MSX 8.3 $\mu$m image from
http://www.ipac.caltech.edu/ipac/msx/msx.html; Price \& Witteborn 1995). The
brightest IR emission is distributed in an arc-like geometry around the R136
cluster core. In the brightest part of this arc, to the NE of R136, is where
the 0539$-$691 maser source is situated. It is also a region of intense
H$\alpha$, [C {\sc ii}] 158 $\mu$m, $^{12}$CO$(1\rightarrow0)$ and far-IR
continuum emission (Poglitsch et al.\ 1995) as well as strong thermal radio
continuum emission (Peck et al.\ 1997). The Mopra beam at 22 GHz covers many
individual IR sources of which some are multiple systems (Walborn et al.\
1999), and in fact there has not yet been identified a definitive counterpart
for the H$_2$O maser in this region. The newly discovered maser source
0539$-$691B is located to the SE of R136, seemingly centred on a moderately
bright blot of IR emission that has received little attention in the
literature and in which no candidate protostars have been proposed.

The H$_2$O masers and associated IR emission seem to be located in between
some of the large expanding shells in the region (Numbers 1 through 5 in Fig.\
2: Cox \& Deharveng 1983; Wang \& Helfand 1991). Some caution should be taken
as to whether these shells are truly expanding or mere superpositions of
non-uniform sheets and filaments (Clayton 1987; Scowen et al.\ 1998). The
masers may be situated near the rim of stellar-wind blown bubbles filled with
hot gas. X-rays trace this hot gas (Wang 1999) and show that bubbles \#3 and
\#5 may already be bursting out of the molecular cloud environment, whilst hot
gas may still be contained within a closed shell \#1.

\section{Discussion}

The 22 GHz detections of 0539$-$691 \& 0539$-$691B show clear morphological
and, for the first time, kinematical evidence for the location of H$_2$O
masers where rapidly expanding volumes of ionized gas collide either with
other such gas volumes or with a neutral and/or molecular medium. The
stellar-wind blown bubbles have supersonically swept up material from the
surrounding ISM, which then became highly compressed and shocked. Upon the
collision of such bubble with another bubble or with a dense cloud, this dense
fast-moving layer becomes even more shocked and compressed. This may be a
favourable condition for the formation of new stars (Dopita 1981; Caulet et
al.\ 1982).

On kinematic grounds we can at least distinguish three masers: two in
0539$-$691 and one in 0539$-$691B. The large velocity difference between the
182 km s$^{-1}$ maser and the other maser emission in 0539$-$691 strongly
suggests also a large spatial separation and the location in distinct
structures in the ISM. The exact location of the H$_2$O masers inside of the
dense molecular layer may be either near the PDR around the protostar and/or
near the PDR associated with the expanding shell. Until interferometric
observations at 22 GHz are undertaken it is not yet possible to identify the
masers with a particular structure as seen at other wavelengths. The reason is
that the morphological and kinematical structure of the ISM within $\sim100$
pc from R136 is extremely complex on all scales.

Protostar candidates in the 30 Doradus complex have been associated with dark
globules that are being photo-evaporated externally (Clayton 1987; Rubio et
al.\ 1992; Scowen et al.\ 1998; Walborn et al.\ 1999) and are situated in
between expanding gas shells (Hyland et al.\ 1992). These shells range in size
from $\sim0.5$ pc around individual massive stars (Hunter et al.\ 1995), up to
the $\sim50$ pc size shells around associations of $\sim10^2$ OB and/or
Wolf-Rayet stars (Chu \& Kennicutt 1994). In fact, both the size distribution
of structures seen in H {\sc i} (Stanimirovi\'{c} et al.\ 1999) and the
size-turbulence relation of H {\sc ii} regions (Roy et al.\ 1986) obey a
Kolmogorov scaling law. The co-existence of ionized and molecular gas in the
PDRs at the interface of the shells and molecular clouds indicates highly
localized (clumpy) and stratified physical structures (Poglitsch et al.\ 1995;
Scowen et al.\ 1998). Poglitsch et al.\ estimate typical clump sizes of a few
pc, containing a mass of $M\sim10^{2-3}$ M$_\odot$. This is just below the
detection limit of the currently available CO surveys. Assuming a star
formation efficiency of $\sim10$\% and a Salpeter-law Initial Mass Function
(Salpeter 1955), the most massive star formed in a $10^3$ M$_\odot$ cloud is
expected to have $M\sim8$ to 10 M$_\odot$ --- i.e.\ an early-B type star with
a luminosity of $L\sim10^4$ L$_\odot$. Hence one or a few bright H$_2$O masers
may be expected in such a star forming cloud.

It seems thus the case that star formation as traced by H$_2$O maser emission
occurs at several locations in the 30 Doradus complex, which are associated
with fast-moving gas ramming into other material. This confirms the picture of
the propagation of sequential star formation, in which stellar-wind blown
bubbles are now starting off a new epoch of star formation in 30 Doradus. It
remains to be proven whether, in a similar fashion, the formation of the young
stellar population of $\sim10^6$ yr --- that includes R136 and that is
responsible for the H {\sc ii} region of 30 Doradus as we see it today --- was
induced by a previous star formation event $\sim2\times10^7$ yr ago as traced
by red supergiants within 45 pc from R136 (Hyland et al.\ 1992).

The star formation propagating throughout the 30 Doradus complex may initially
have been triggered by ram pressure resulting from the supersonic motion of
the LMC through the extended galactic halo (de Boer et al.\ 1998): the LMC
appears to have been moving from apo-galacton at $\sim100$ to 200 kpc one Gyr
ago to peri-galacton at $\sim50$ kpc today (Gardiner et al.\ 1994; Heller \&
Rohlfs 1994). A close encounter between the Magellanic Clouds $\sim2$ to
$5\times10^8$ yr ago (Gardiner et al.\ 1994; Heller \& Rohlfs 1994) may also
have played a r\^{o}le in stirring up the ISM in the LMC, triggering enhanced
star formation as evidenced by common peaks in the age distributions of
clusters in both Magellanic Clouds (Pietrzy\'{n}ski \& Udalski 2000). More may
be learnt about the triggering of sequential star formation from studies of
the 3 kpc long reservoir of CO gas due S of 30 Doradus (Israel 1984; Cohen et
al.\ 1988; Kutner et al.\ 1997) that seems to be in an earlier stage of star
formation activity than the 30 Doradus complex (Heydari-Malayeri et al.\
1999).

If the newborn stars have space velocities similar to those of the H$_2$O
masers, then these stars will quickly mix with the older stellar populations
of the LMC. Together with the ejection of stars by supernovae in binaries
(Zwicky 1957; Blaauw 1961; Kaper et al.\ 1997; Hoogewerf et al.\ 2000) and
dynamical interactions within young clusters (Poveda et al.\ 1967; Gies \&
Bolton 1986; Hoogewerf et al.\ 2000), high-velocity star formation may thus
provide an additional mechanism for the formation of runaway stars. In fact,
it has been argued that the star cluster associated with SN 1987A has formed
at a velocity of 50 km s$^{-1}$ relative to the disk (Graff \& Gould 2000).
Such high velocity corresponds to 5 kpc in $10^8$ yr, sufficient to cross the
bar of the LMC. Thus both gas and young stars may efficiently mix throughout
the entire galaxy.

There is some indication that this may also happen in the Milky Way: the
presence of early-type Main-Sequence stars in the galactic halo (Rolleston et
al.\ 1999) may be explained by their formation in gas entities that were being
ejected out of the galactic disk. Some of this gas may fall back towards the
galactic disk in the form of high-velocity clouds (Richter et al.\ 1999). We
see the galactic disk edge-on, but the LMC disk nearly face-on, which makes it
much easier to detect this escaping gas and associated star formation in the
LMC than in our own Milky Way galaxy.

\acknowledgments

We would like to thank Gerry Gilmore, Tom Millar, Malcolm Gray and Gary Fuller
for useful discussions, Robina Otrupcek for support during the observations at
Mopra, and the referee for her/his remarks. Jacco thanks Joana for being who
she is.

\clearpage

\figcaption[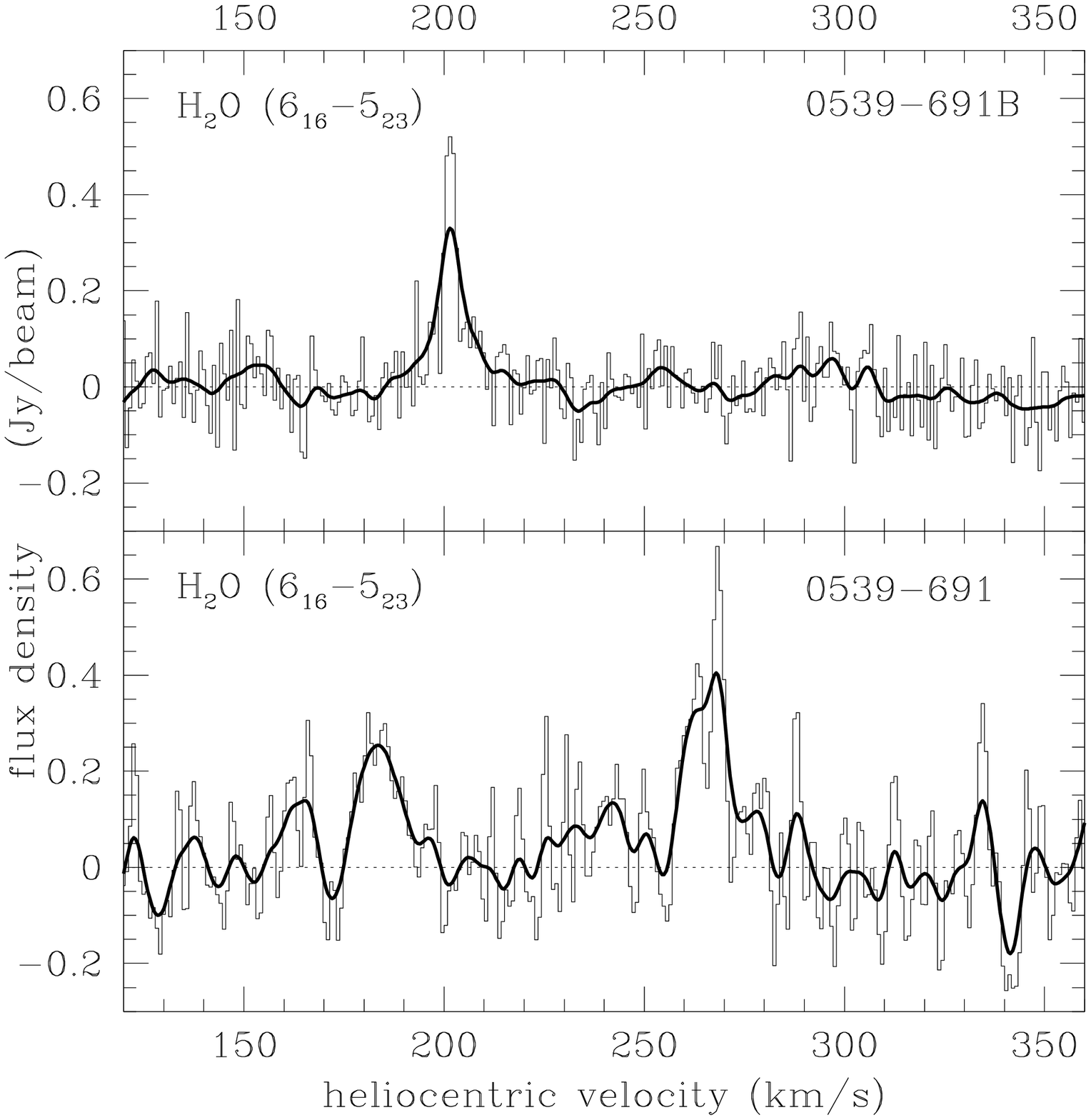]{H$_2$O maser emission (22 GHz, Mopra) from 0539$-$691B
(top panel) and 0539$-$691 (bottom panel). The velocities are heliocentric.
The boldfaced curves are the spectra smoothed by a gaussian of $\sigma=2$ km
s$^{-1}$. \label{fig1}}

\figcaption[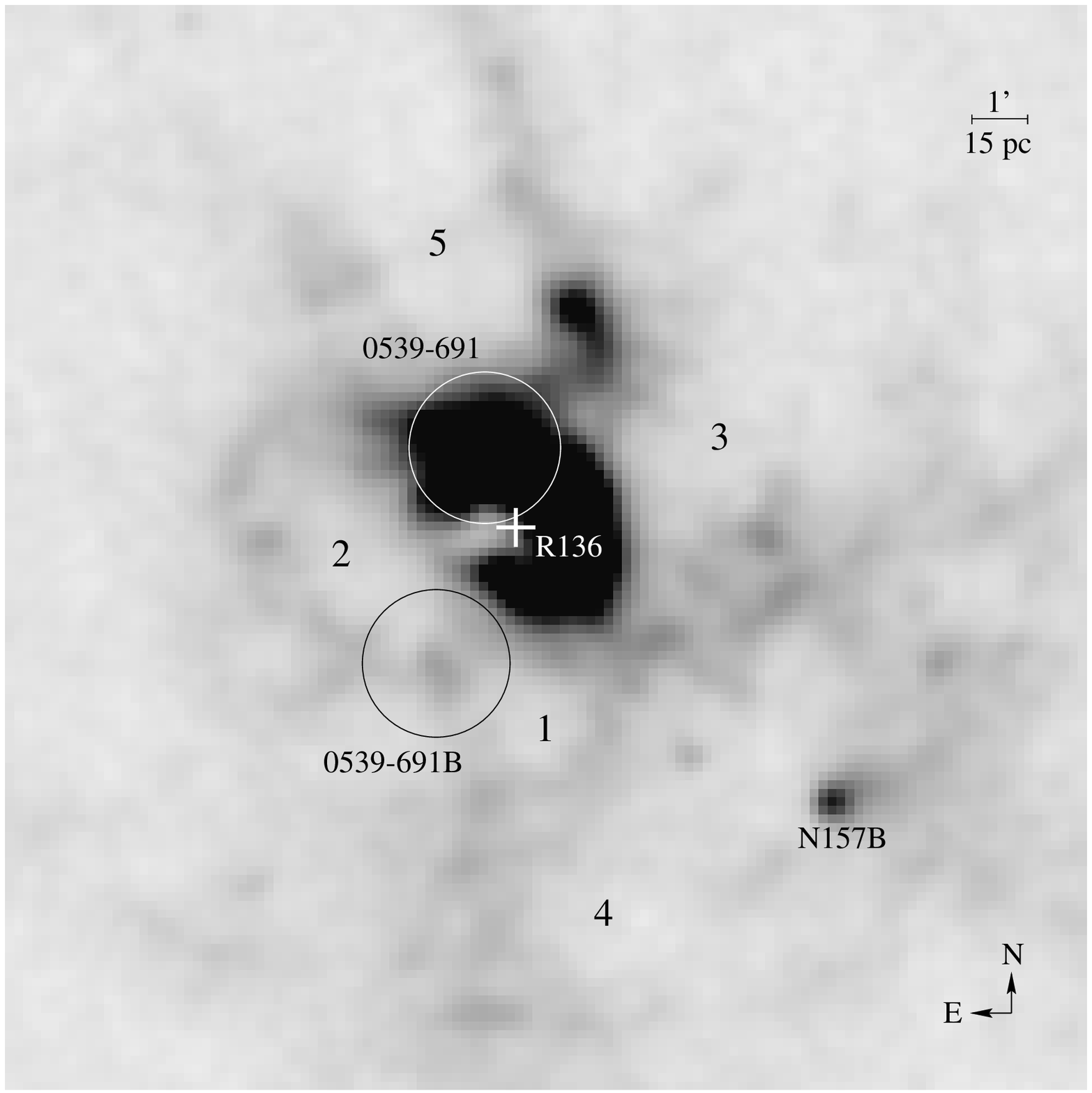]{MSX 8.3 $\mu$m image of the 30 Doradus region, with
the beams pointed at the H$_2$O maser sources 0539$-$691 and 0539$-$691B, as
well as the location of the OB association R136, the supernova remnant N157B,
and five shells as designated by Cox \& Deharveng (1983) and Wang \& Helfand
(1991). \label{fig2}}

\end{document}